\documentclass[12pt,preprint]{aastex}
\usepackage{natbib}

\slugcomment{Accepted to AJ, 2008 Nov 28}

\shorttitle{Young L Dwarfs in the Field}

\shortauthors{Cruz, Kirkpatrick, \& Burgasser}

\begin{document}

\title{Young L Dwarfs Identified in the Field: A Preliminary Low-Gravity, Optical Spectral Sequence from L0 to L5.}

\author{Kelle L. Cruz\altaffilmark{1}}
\affil{Astronomy Department, California Institute of Technology, Pasadena, CA 91125, USA} 
\email{kelle@astro.caltech.edu}

\author{J. Davy Kirkpatrick}
\affil{Infrared Processing and Analysis Center, California Institute of Technology, Pasadena, CA 91125, USA}

\and

\author{Adam J. Burgasser}
\affil{Kavli Institute for Astrophysics and Space Research, Massachusetts Institute of Technology, Cambridge, MA 02139, USA}

\altaffiltext{1}{Spitzer Postdoctoral Fellow}

\begin{abstract}

We present an analysis of 23 L dwarfs whose optical spectra display unusual features.
Twenty-one were uncovered during our search for nearby, late-type objects using the Two Micron All-Sky Survey while two were identified in the literature.
The unusual spectral features, notably weak FeH molecular absorption and weak \ion{Na}{1} and \ion{K}{1} doublets, are attributable to low-gravity and indicate that these L dwarfs are young, low-mass brown dwarfs.
We use these data to expand the spectral classification scheme for L0 to L5-type dwarfs to include three gravity classes.
Most of the low-gravity L dwarfs have southerly declinations and distance estimates within 60~pc.
Their implied youth, on-sky distribution, and distances suggest that they are members of nearby, intermediate-age ($\sim$10--100 Myr), loose associations such as the $\beta$~Pictoris moving group, the Tucana/Horologium association, and the AB Doradus moving group.
At an age of 30~Myr and with effective temperatures from 1500 to 2400~K, evolutionary models predict masses of 11--30~$M_{Jupiter}$ for these objects.
One object, 2M~0355+11, with $J-K_s=2.52\pm0.03$, is the reddest L dwarf found in the field and its late spectral type and spectral features indicative of a very low gravity suggest it might also be the lowest-mass field L dwarf.
However, before ages and masses can be confidently adopted for any of these low-gravity L dwarfs, additional kinematic observations are needed to confirm cluster membership.

\end{abstract}


\section{Introduction}

Brown dwarfs are star-like objects with insufficient mass to sustain hydrogen burning ($M\la75~M_{Jupiter}$). 
As a result, they never reach the main sequence and instead, continually cool with time and evolve through the MLT spectral sequence \cite[and references therein]{Kirkpatrick05}. 
Unlike main sequence stars, the spectral type of a field brown dwarf does not provide a unique constraint on its mass.
For example, an early-L dwarf could be an old low-mass star, a young low-mass brown dwarf, or an even younger planetary-mass brown dwarf. 

Observations of spectral features that are sensitive to surface-gravity effects ($g \propto M/R^2$) have long been used to distinguish compact dwarf stars from massive, extended giants with similar effective temperatures.
The recognition of similar gravity-sensitive features in the spectra of brown dwarfs has begun to help break the age-mass degeneracy that has complicated their study.
Brown dwarfs younger than $\sim$100~Myr have low gravities because they have both larger radii (since they are still contracting) and lower masses than their same spectral type counterparts in the field \citep{Burrows01}. 
Spectral features indicative of low-gravity (weak CaH, \ion{K}{1}, \ion{Na}{1}, and strong VO) have been identified in the optical spectra of late-M dwarf members of young 1--5~Myr clusters 
\citetext{e.g., IC 348, \citealp{Luhman99_IC348}; Taurus, \citealp{Luhman03_taurus}}, the juvenile $\sim$10~Myr TW Hydrae association \citep{Gizis02}, and the adolescent 100~Myr Pleiades open cluster \citep[e.g.,][]{Martin96}. 

While optical (6000--9000~\AA) spectra of very young M dwarfs are relatively common, similar data for L dwarfs, which have extremely faint optical magnitudes, are sparse.
That said, optical spectroscopy has revealed several candidate members of $\sigma$~Orionis, Lupus, Chamaeleon I, and Chamaeleon II to have early-L spectral types \citep{BarradoyNavascues01,Jayawardhana06,Luhman08a,Luhman08b}.
However, the low resolution and low signal-to-noise of those data preclude any examination of the low-gravity features that might be present.
Near-infrared (1--2.5~\micron) spectra of the faintest young cluster brown dwarfs are more feasible to obtain and have revealed candidate L dwarf members of Orion, Chameleon II, Ophiuchus, and Upper Scorpius \citep{Lucas01, McGovern04, Allers07,lodieu08_spectra}.
Unfortunately, near-infrared data of young objects systematically yield later spectral types than optical spectra \citep{Luhman03_IC348}. 
For example, \citet{Herczeg08_lris} find that several of the latest L dwarf candidates identified in Upper Sco with near-infrared spectra by \citet{lodieu08_spectra} are actually late-M dwarfs in the optical.
In short, unlike for M dwarfs, there are presently no optical spectra of 1--10~Myr-old L dwarfs with sufficient resolution or signal-to-noise ratios to serve as age benchmarks for L dwarfs found in the field.

There are two well-studied L dwarfs outside young clusters that display unambiguous low-gravity features in their optical and near-infrared spectra: G196-3B and \objectname[2MASS J01415823-4633574]{2MASS J01415823$-$4633574} (hereafter 2M~0141$-$46
\footnote{We use abbreviated notation for 2MASS sources throughout the text; e.g., 2M~hhmm$\pm$dd, where the suffix is the J2000 sexagesimal right ascension (hours and minutes) and declination (degrees). Full source names and coordinates are provided in Table~\ref{tab:lowg}.}). 
G196-3B was identified as an L2-type companion to the young ($\sim$20--300~Myr) early-M dwarf G196-3A \citep{Rebolo98, K01_gstar} while 2M~0141$-$46 was discovered as a single L0-type dwarf in the field \citep{Kirkpatrick06}. 
\citet{McGovern04} performed a thorough optical and near-infrared spectral analysis of G196-3B while this was done for 2M~0141$-$46 in its discovery paper.
These authors find that G196-3B and 2M~0141$-$46 resemble early-type L dwarfs but have peculiar optical and near-infrared spectral features indicative of low surface-gravity ($log(g) \sim~4.0 \pm 0.5$).
In particular, the spectra display weak \ion{Na}{1} doublets (8183, 8195~\AA; 1.13, 1.14~\micron), weak and sharp \ion{K}{1} doublets (7665, 7699~\AA; 1.17, 1.24~\micron), strong VO absorption bands (7300--7550, 7850--8000~\AA; 1.05~\micron), and weak FeH absorption bands (0.98, 1.19~\micron).
Based on the age estimate for G196-3A, \citeauthor{Rebolo98} estimate the mass of G196-3B to be $25_{-10}^{+15} M_{Jupiter}$.
Comparing the spectra of 2M~0141$-$46 to both models and late-type dwarfs with well-constrained ages, \citeauthor{Kirkpatrick06} estimate 2M~0141$-$46 to have an age of 1--50~Myr and a mass of 6--25~$M_{Jupiter}$.
Given its estimated age and location deep in the southern hemisphere, \citeauthor{Kirkpatrick06} also hypothesize that 2M~0141$-$46 might be a member of the 12~Myr-old $\beta$~Pictoris moving group or the 30~Myr-old Tucana/Horologium association.

In addition to these two benchmark objects, other L dwarfs with low-gravity features in their optical spectra have been identified in the field.
Three were presented by \citet{Cruz07} and fourteen were listed by \citet{Paper10}.
Combining these discoveries with new ones and improved data, \citet{Kirkpatrick08} examined in detail the optical spectra of twenty low-gravity objects, including eight late-M dwarfs and eleven L dwarfs. 
In this paper, we focus only on L dwarfs, complementing the efforts of \citeauthor{Kirkpatrick08}. 
Here we identify and examine twenty-three low-gravity L dwarfs found in the field and use extant data to compile a spectral sequence that spans from L0 to L5 and includes three gravity classes.
These data include eight very low-gravity L0-type objects with spectra nearly identical to that of 2M~0141$-$46; two additional L0 dwarfs with spectral features suggesting intermediate gravities; and thirteen objects, including G196-3B, that have low gravities and even later spectral types (L1--L5).

In \S~\ref{sec:id} we briefly outline how the low-gravity objects were identified during the course of the NStars census of late-type dwarfs within 20~pc of the Sun.
In \S~\ref{sec:types} we introduce a spectral typing scheme for low-gravity L dwarfs.
In \S~\ref{sec:spectra}, the unusual spectra are described and compared to spectral standards via both overplotting and spectral indices. 
We use empirical arguments to justify low-gravity as the explanation for the observed spectral peculiarities.
In \S~\ref{sec:discussion}, we discuss age, distance, and mass estimates for the low-gravity objects and, based on their distribution on the sky, suggest that they are likely members of nearby young moving groups. 
We summarize our results in \S~\ref{sec:summary}.

\section{Identification of Low-Gravity Candidates}
\label{sec:id}

In Table~\ref{tab:lowg} we list twenty-three L dwarfs that exhibit spectral features indicative of low gravity. 
Most of these objects (21/23) were found during the course of the 20-pc census for nearby brown dwarfs using 2MASS; the selection criteria used to identify them are described extensively in other papers \citep{Cruz03, Paper10}. 
Two objects included in our analysis were identified in the literature and not found by the 20-pc census: 2M~2208+29 and G196-3B.
2M~2208+29 was discovered by \citet{K00} where it was noted as having a peculiar spectrum.
G196-3B was discovered by \cite{Rebolo98} during a direct-imaging search around nearby, young K and M dwarfs.
Three low-gravity L dwarfs found during the 20-pc census were also independently identified in the Southern Sample of \cite{Kirkpatrick08}: 2M~0033$-$15, 2M~0141$-$46, and 2M~0357$-$44 \citep[aka DENIS 0357$-$44, ][]{Bouy03}.
(Including these three, there are seven L dwarfs in common between the twenty-three L dwarfs considered here and the twenty objects studied by \citet{Kirkpatrick08}.)

The primary criteria used to identify candidate nearby brown dwarfs for the 20-pc census were two near-infrared color-cuts:
\begin{displaymath}
(J-K_s) > 1.0 
\end{displaymath}
for the 2MASS Second Incremental Data Release Point Source Catalog,
\begin{displaymath}
(J-K_s) > 1.06
\end{displaymath}
for the 2MASS All-Sky Point Source Catalog, and
\begin{displaymath}
J \le 3(J-K_s) + 10.5
\end{displaymath}
for both catalogs\footnote{The slope of this equation was incorrectly stated as 1.5 in \cite{Cruz03}. 
The correct value is 3 and the line is plotted correctly in Fig. 1 of that paper.}. 
Further details about the creation of the 20-pc samples are given in \citet{Cruz03} and \citet{Paper10}. 
The identification of low-gravity brown dwarfs was not considered when creating these selection criteria. 

%495 + 376
Several different telescopes were used to obtain moderate resolution ($R\sim1000$) optical spectra (6000--9000~\AA) of $\sim$900 objects that survived the selection criteria for the 20-pc census. 
The spectrum of one object, 2M~0712$-$61, was obtained recently and is the only new discovery listed in this paper.
The observation date, telescope, and instrument used to obtain the peculiar spectra discussed here are listed in Table~\ref{tab:lowg}; the instrumental setup and data reduction techniques used are described in the papers referenced in the table.
Every spectrum was visually compared to spectral standards in order to estimate spectral types. 
Through this process, a handful of peculiar M and L dwarfs were identified that did not match any standard and also displayed spectral features indicating low-gravity. 
These peculiar M and L dwarfs were flagged for further analysis and noted as low gravity in their discovery papers \citep{Cruz03,Cruz07,Paper10}.
We have found other L-type objects with features indicative of low gravity (e.g., 2MASS~1615+49, \citealt{Cruz07,Kirkpatrick08}) but chose not to include them here pending higher signal-to-noise ratio spectra necessary to verify the low-gravity features.
Hence, the sample presented here should not be considered complete.

\section{A Spectral Typing Scheme for Low-Gravity L Dwarfs}
\label{sec:types}

\begin{figure}
\plotone{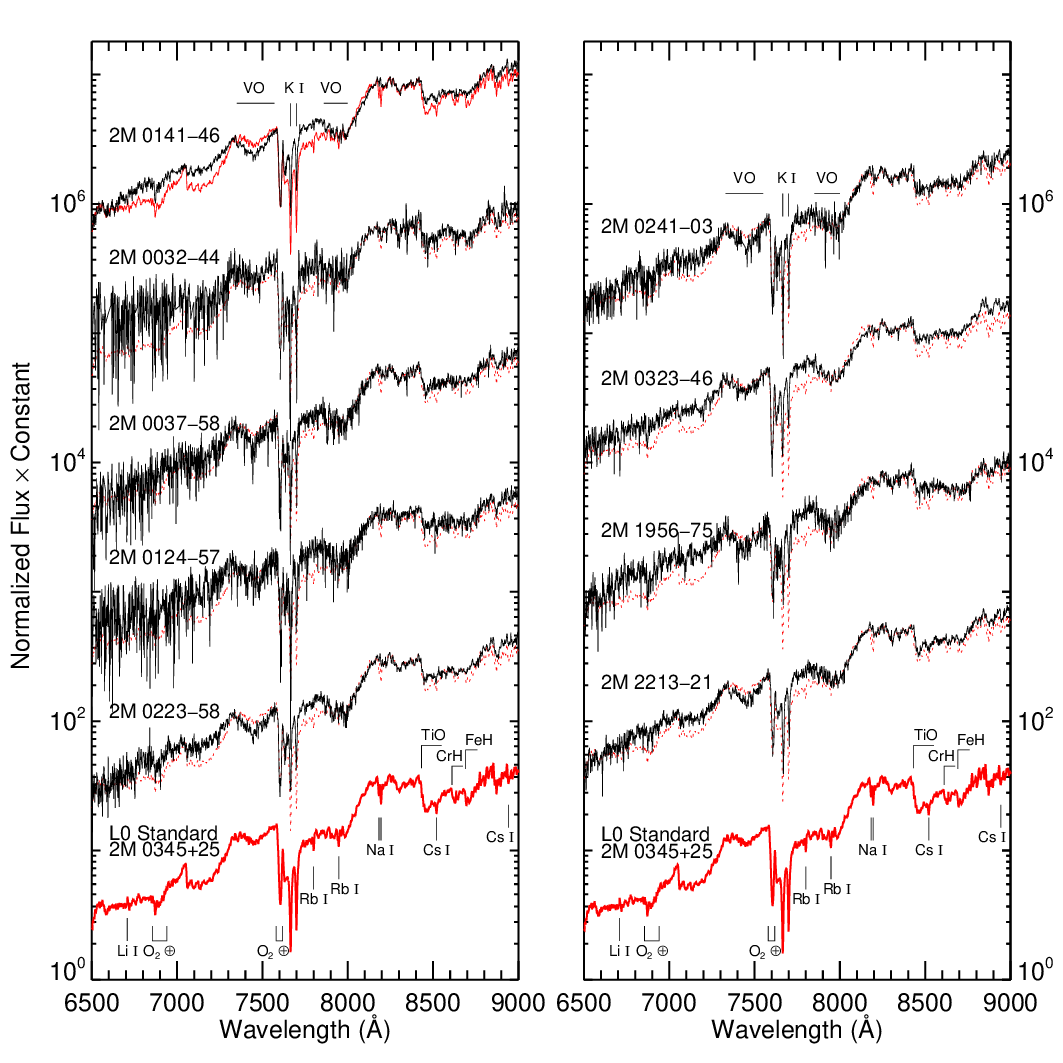}
\caption[2M~0141$-$46 clones] {
	Red-optical spectra of L0$\gamma$-type dwarfs that exhibit spectral features indicative of very low-gravity (\textit{black}). 
	All of these objects are near clones of 2M~0141$-$46 \citep{Kirkpatrick06}, which is shown at top and compared to the L0 standard 2M~0345+25 (\textit{thick red}). 
	The L0 standard is also overplotted on the clones (\textit{red dotted}) and shown at the bottom of each panel (\textit{thick red}). 
	None of these data are telluric corrected. 	
	All data are normalized at 8240--8260~\AA.
	Gravity-sensitive spectral features are labeled. 
	The y-scale is logarithmic and the y-range is not the same amongst the spectra figures.}
\label{fig:clones}
\end{figure}

\begin{figure}
\plotone{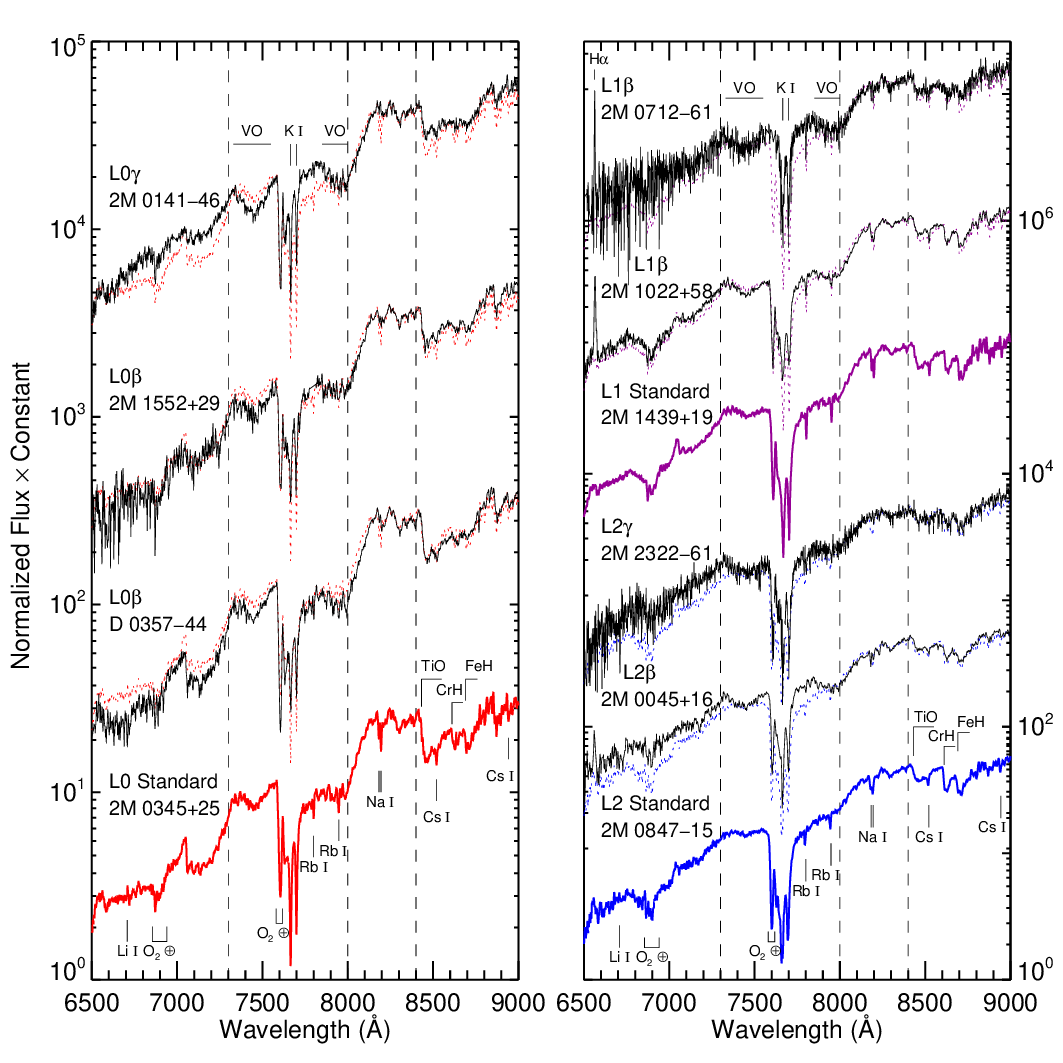}
\caption[spectral sequence of low-gravity L dwarfs]{
	Red-optical spectral sequence of low-gravity L0--L2 dwarfs (\textit{black}, $\gamma$: very low-gravity, $\beta$: intermediate low-gravity) and normal-gravity spectral standards (L0: \textit{red}, L1: \textit{purple}, L2: \textit{blue}).
	Within each sub-type, objects are plotted from top to bottom in decreasing order of the prominence of their low-gravity features with the normal-gravity spectral standard shown last (\textit{thick solid}). 
	The spectral standard is also overplotted on the low-gravity spectra (\textit{dotted}). 
	The spectrum of 2M~0712$-$61 (L1$\beta$) has been corrected for telluric absorption.
	All data are normalized at 8240--8260~\AA. 
	Gravity-sensitive spectral features are labeled.
	The wavelength region most affected by gravity (7300--8000~\AA) and the relatively gravity-insensitive neighboring region (8000--8400~\AA) are demarcated by dashed lines.
	The y-scale is logarithmic and the y-range is not the same from panel to panel.}
\label{fig:sequence1}
\end{figure}

\begin{figure}
\epsscale{1}
\plotone{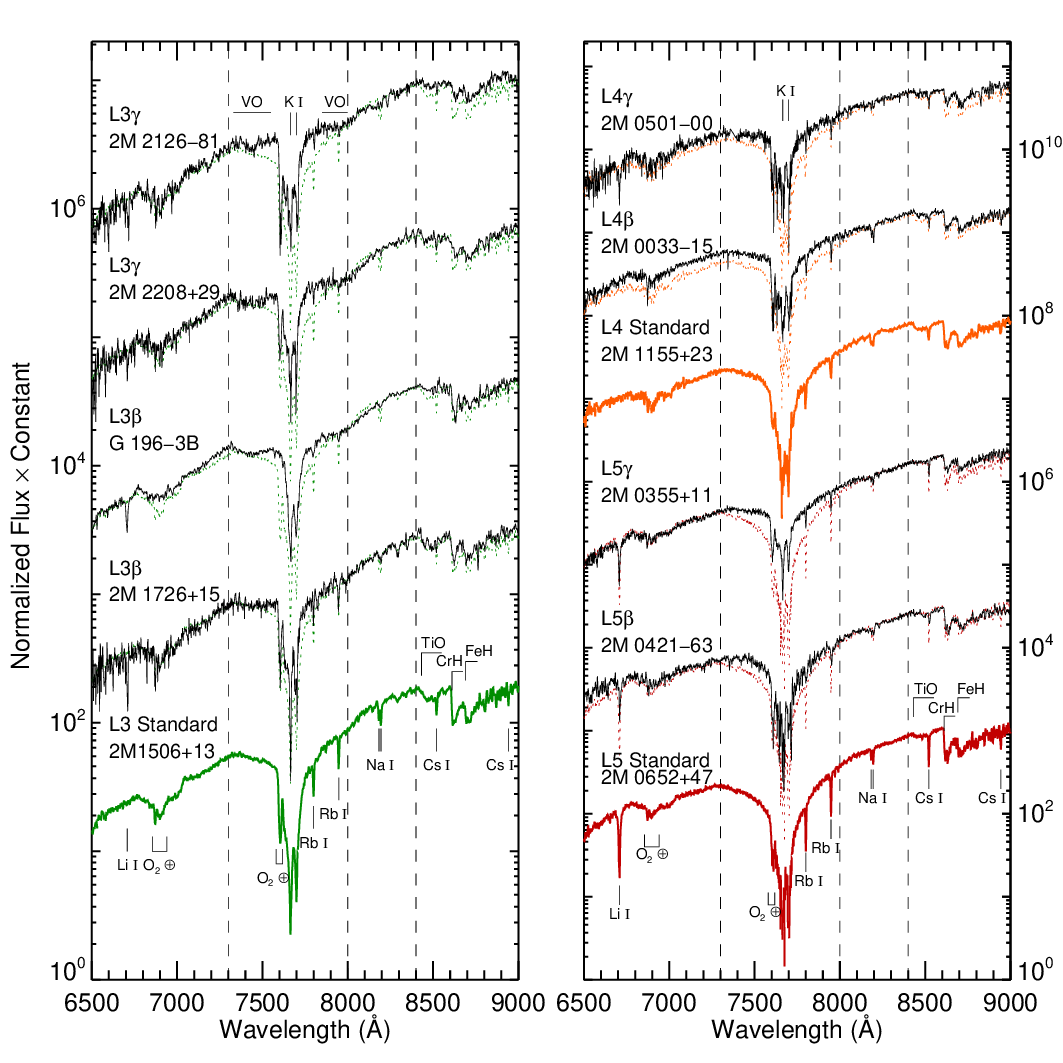}
\caption[spectral sequence of low-gravity L dwarfs] {
	Red-optical spectral sequence of low-gravity L3--L5 dwarfs (\textit{black}, $\gamma$: very low-gravity, $\beta$: intermediate low-gravity) and normal-gravity spectral standards (L3: \textit{green}, L4: \textit{orange}, L5: \textit{red}). 
	Within each sub-type, objects are plotted from top to bottom in decreasing order of the prominence of their low-gravity features with the normal-gravity spectral standard shown last (\textit{thick solid}). 
	The spectral standard is also overplotted on the low-gravity spectra (\textit{dotted}).
	The spectrum of G196-3B (L3$\beta$) has been corrected for telluric absorption. 
	All data are normalized at 8240--8260~\AA. 
	Gravity-sensitive spectral features are labeled. 
	The wavelength region most affected by gravity (7300--8000~\AA) and the relatively gravity-insensitive neighboring region (8000--8400~\AA) are demarcated by dashed lines.	
	The y-scale is logarithmic and the y-range is not the same from panel to panel.}
\label{fig:sequence2}
\end{figure}

The spectra shown in Figures~\ref{fig:clones}--\ref{fig:sequence2} significantly deviate from the spectral standards but most resemble L dwarfs.
As described in the following section, the spectral morphology of these objects and the standards deviate most where the gravity-sensitive pressure-broadened \ion{K}{1} absorption doublet shapes the spectrum. 
However, the general agreement of the peculiar spectra with the normal-gravity standards is used to classify them as early-to-mid L dwarfs (as opposed to M or T dwarfs).
In addition to the entire spectral shape, the spectral region from $\sim$8000--8400~\AA, excluding the \ion{Na}{1} doublet, is used to determine the spectral subtype via comparison to the \citet{K99} L dwarf spectral standards and the supplementary standards listed by \citet{Paper10}.
In this wavelength regime, the low-gravity spectra match the standards and there is a marked change in the spectral shape from one subtype to the next: the plateau present in the early-L types gradually evolves to a smooth red slope in the mid-L types.
Both this region and the portion of the spectrum most affected by the pressure-broadened \ion{K}{1} absorption are marked on Figs.~\ref{fig:sequence1} and~\ref{fig:sequence2} by dashed lines.
On this spectral typing scheme, it is likely that the only physical characteristic shared by objects of the same spectral subtype is spectral morphology in the 8000--8400~\AA\ region---objects of the same subtype with different gravities probably span a range of effective temperatures and certainly have different ages and masses.

Within each subtype, we have used the weakness of the \ion{Na}{1} doublet and the \ion{K}{1} doublet cores and wings to distinguish objects with not-as-prominent low-gravity features from those with very conspicuous low-gravity features.
As suggested by \citet{Kirkpatrick05} and further discussed by \cite{Kirkpatrick06}, we indicate intermediate-gravity with a $\beta$ appended to the spectral type and very low-gravity spectra with a $\gamma$.
An $\alpha$ suffix is implied for normal-gravity objects.
This greek suffix notation replaces the two different conventions that have been used to indicate low-gravity:  ``pec'' suffix \citep[indicating peculiar, ][]{Kirkpatrick06,Kirkpatrick08} and enclosing the spectral type in parentheses \citep{Cruz07,Paper10}.

\defcitealias{K99}{K99}

The \citet[hereafter K99]{K99} scheme was created to type normal L dwarfs and, understandably, does not take gravity effects into account.
The low-gravity scheme introduced here is based on \citetalias{K99} and should be considered an expansion of that scheme.
For dwarfs of normal gravity, the \citetalias{K99} scheme needs no modification.
This new low-gravity scheme is still preliminary; more sources and higher signal-to-noise data will eventually enable a formal low-gravity L dwarf classification scheme to be developed with spectral standards and guidelines based on spectral indices.
Concurrent work is underway to create a low-gravity sequence for M dwarfs that is tied to young cluster members with known ages (Kirkpatrick et al., in preparation).

\cite{Kirkpatrick08} use the \citetalias{K99} recipe to assign spectral types to low-gravity objects and as a result, there is some disagreement between the spectral types of objects in both samples.
In particular, the typing scheme used here and the \citetalias{K99} system agree at early types (L0--L1) but diverge at later types: \citet{Kirkpatrick08} find 2M~0033$-$15, 2M~2208+29 and G~196-3B to be L2 type dwarfs while here they are typed as L4, L3, and L3, respectively.

\section{Spectral Properties of Candidate Low-Gravity L Dwarfs}
\label{sec:spectra}

In Figure~\ref{fig:clones} we show nine very low-gravity L0$\gamma$-type dwarfs, including 2M~0141$-$46 \citep{Kirkpatrick06}.  
In Figure~\ref{fig:sequence1}, we show 2M~0141$-$46 and six additional low-gravity L0--L2 type dwarfs. 
Eight low-gravity L3--L5 type dwarfs are shown in Figure~\ref{fig:sequence2}, including G196-3B. 
In the latter two figures, the spectra within the same subtype are plotted from top to bottom in approximate decreasing order of their spectral peculiarity with the normal-gravity standard shown last.

\begin{figure}
\plotone{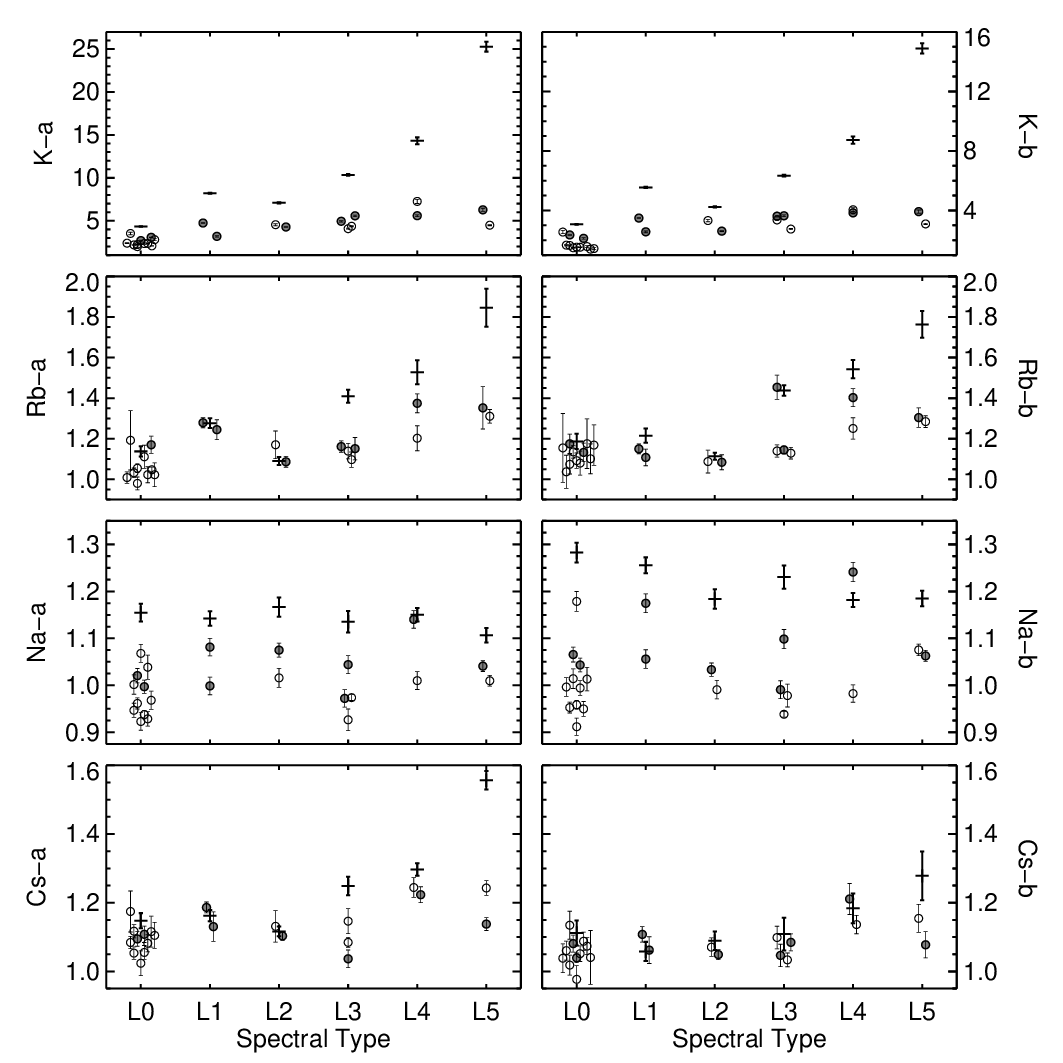}
\caption[Indices of the alkali metal lines]{
	Alkali-metal spectral-index values as a function of spectral type.
	The K-a and K-b indices are defined in \S~\ref{sec:spectra} while the other indices are defined by \cite{K99}. 
	The index values for the very low-gravity (\textit{open circles}) and intermediate gravity (\textit{shaded circles}) objects are compared to the values for the normal-gravity standards (\textit{dashes}).
	Overlapping data points are offset along the x-axis for clarity.}
\label{fig:lines}
\end{figure}

\begin{figure}
\plotone{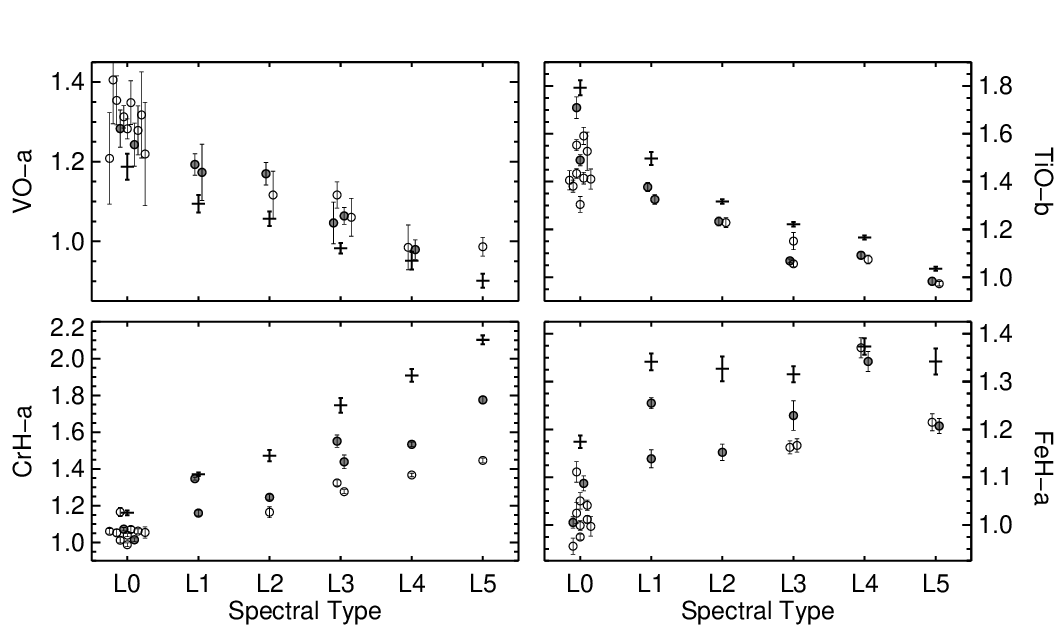}
\caption{
	Molecular spectral-index values as a function of spectral type.
	These indices are defined by \cite{K99}.
	The index values for the very low-gravity (\textit{open circles}) and intermediate gravity (\textit{shaded circles}) objects are compared to the values for the normal-gravity standards (\textit{dashes}).
	Overlapping data points are offset along the x-axis for clarity.}
\label{fig:bands}
\end{figure}

In Figures~\ref{fig:lines} and~\ref{fig:bands}, the values of twelve spectral indices measuring gravity-sensitive features for each low-gravity object and normal-gravity standard are shown. 
Ten of the indices were defined by \citet{K99} and we introduce two new ones that measure the depths of the \ion{K}{1} lines at 7665 and 7699~\AA:
\begin{displaymath}
\mbox{K-a} = \frac{\langle F_{\lambda=7550-7570}\rangle}{\langle F_{\lambda=7655-7675}\rangle} \mbox{ and}
\end{displaymath}
\begin{displaymath}
\mbox{K-b}=\frac{\langle F_{\lambda=7550-7570}\rangle}{\langle F_{\lambda=7690-7710}\rangle}.
\end{displaymath}
Uncertainties on all of the indices were estimated using the standard deviation of the mean flux ($\sigma$) in the region of pseudo-continuum used by the index (i.e., the numerator).
The region used to estimate the uncertainty is centered on the same wavelength as the numerator, but twice as many Angstroms wide.
The resulting percent uncertainty ($\sigma$/mean flux) is then adopted as the uncertainty for both the index numerator and denominator.
The $\sigma$ of the denominator is not used since it reflects the shape of the feature being measured rather than the signal-to-noise ratio of the pseudo-continuum.
There are six spectral index measurements where noise spikes alter the index value so significantly that it does not reflect the true depth of the feature being measured.
These points are omitted from Figs.~\ref{fig:lines} and~\ref{fig:bands} and include VO-a of 2M~0421$-63$, Cs-a and FeH-a of G196-3B, Rb-a of 2M~1552+29, TiO-b of 2M~1726+15, and FeH-a of 2M~2322$-61$.

There are several atomic and molecular features that distinguish the low-gravity objects from normal L dwarfs.
Below, we describe the unusual features observed in the spectra and discuss the empirical evidence pointing to low-gravity as their cause. 
We leave a discussion of the physical explanation for the gravity-sensitive spectral features to a future paper  (Marley et al., in preparation).

\subsection{Alkali Metals: K, Na, Rb, \& Cs}

The most prominent feature in low-gravity L dwarfs is weaker absorption due to the \ion{K}{1} doublet (7665, 7699~\AA) and its pressure-broadened wings. 
In the low-gravity early-Ls (L0--L2) the doublet lines are both weaker and sharper than in normal dwarfs (Fig.~\ref{fig:sequence1}).
In addition, the later-type (L3--L5) low-gravity L dwarfs have more flux than normal dwarfs in the 700~\AA\ surrounding the doublet (Fig.~\ref{fig:sequence2}, left region between dashed lines).
Absorption due to the other alkali metals is also weaker in the low-gravity objects than in normal dwarfs.
In particular, the \ion{Na}{1} doublet (8183, 8195~\AA) in most objects is either not present or very weak and the \ion{Rb}{1} (7800, 7948~\AA) and \ion{Cs}{1} (8521, 8943~\AA) lines are weak in the later type (L2--L5) objects. 
The weakening of \ion{Na}{1} and \ion{K}{1} doublets at low gravities has been long-established empirically: these lines are not present in the spectra of late-type giants ($log (g) \sim (0)$) and weak \ion{Na}{1} and \ion{K}{1} absorption are commonly seen in late-type M dwarf members of young clusters \citep[e.g.,][]{Martin96,Luhman97,Slesnick04}.
Weak \ion{Rb}{1} and \ion{Cs}{1} lines in conjunction with weak \ion{Na}{1} and \ion{K}{1} absorption is observed in the 20--300~Myr L dwarf \object{G~196-3B} \citep{McGovern04}.

In Figure~\ref{fig:lines} we compare the values of the \citetalias{K99} alkali-line spectral indices and the new \ion{K}{1} indices measured for the very low-gravity (\textit{open circles}), the intermediate-gravity (\textit{shaded circles}) objects, and the spectral standards (\textit{dashes}).
These indices were not used to distinguish the very low-gravity and intermediate-gravity objects.
In general, these plots quantify and confirm our by-eye observation of weaker lines in the low-gravity objects compared to the normal-gravity standards.
The indices measuring the \ion{K}{1} and \ion{Na}{1} doublets, in particular, consistently have lower values for the low-gravity objects than for the standards.
However, the inherent weakness of the \ion{Cs}{1} and \ion{Rb}{1} lines at L0--L2 spectral types, combined with the relatively low signal-to-noise of our data, results in index values that are close to unity and that do not clearly discriminate between the low-gravity objects and the standards.
Indeed, as can be seen in the spectra, these lines are barely present, especially at early types, in both the low-gravity spectra and the standards.
Thus, alkali-line spectral indices cannot (and should not) be used alone to quantitively define a low-gravity spectrum.
Instead, low gravity should be diagnosed with by-eye comparative analysis in conjunction with spectral indices measuring both line and band strengths.

\subsection{Metal-Oxides and -Hydrides: VO, TiO, CrH, \& FeH}
Several metal-oxide and metal-hydride absorption features have different strengths in the low-gravity L dwarfs than in normal-gravity dwarfs.
In the early-type (L0--L3) low-gravity objects, one of the most distinguishing features is the deep molecular absorption bands of VO at 7300--7550~\AA\ and 7850--8000~\AA.
These bands appear much stronger than in normal dwarfs and remain present to L3 in the low-gravity sequence. 
In the lowest-gravity objects of all subtypes, there are also slightly weaker molecular absorption bands of TiO (8432~\AA), CrH (8611~\AA), and FeH (8692~\AA).

A close inspection of Figure~\ref{fig:sequence1} reveals that lowest-gravity early-type objects appear to have more flux in the pseudo-continuum between the \ion{K}{1} lines and the 7850--8000~\AA\ VO band causing the bottom of the VO band in the low-gravity objects to be at the same depth as the normal-gravity objects.
We have investigated the behavior of this spectral region in giants and dwarfs of both early and later spectral types and conclude that this pseudo-continuum bump is an artifact of the somewhat arbitrary normalization point and the difference in the relative flux levels in objects of different gravities.
Based on the similarity of the VO band shapes to those seen in giants, we are confident that the spectral shape of the peculiar objects in this region is primarily a result of \textit{stronger} VO absorption.

Similar to \ion{Na}{1} and \ion{K}{1}, the behavior of VO and FeH absorption bands at low gravities has been established empirically:
In both late-M giants and young dwarfs, the VO bands at 7300--7550~\AA\ and 7850--8000~\AA\ are strong while the FeH band at 1.2~\micron\ is weak \citep{Martin96,Gizis02,Briceno02,McGovern04}.
However, contrary to our observation of weaker TiO, TiO band strengths have been found to be stronger in late-M giants and young dwarfs than in normal-gravity objects \citep{McGovern04}.
Weak TiO absorption is the only inconsistency between the spectral peculiarities observed in our candidate low-gravity L dwarfs and those observed in low-gravity fiducials.
Parallaxes might help resolve this TiO issue by ensuring that the objects with different gravities that we are comparing to each other also have similar effective temperatures.

In Figure~\ref{fig:bands}, we compare the values of the \citetalias{K99} molecular bands indices measured for the very low-gravity (\textit{open circles}), the intermediate-gravity (\textit{shaded circles}) objects, and the spectral standards (\textit{dashes}).
Again, these indices were not used to distinguish the very low-gravity and intermediate-gravity objects.
Overall, these molecular indices distinguish the low-gravity objects from the normal-gravity standards.
However, we note that, except for FeH, the separation of the low-gravity objects from the standards would not be as distinct if the low-gravity objects were typed one class earlier (for VO and CrH) or later (for TiO).
Thus, we emphasize that careful visual inspection of the entirety of the spectrum is required to diagnose low-gravity and strongly caution against using spectral indices alone for spectral classification.

\subsection{Lithium}

The \ion{Li}{1} unresolved absorption doublet at 6708~\AA\ is seen mostly in the later-type low-gravity objects (L3--L5) and not in the earlier ones, as is also the case for normal L dwarfs.
The equivalent widths and upper limits on the detection of this line are listed in Table~\ref{tab:lowg}.
We measured these values by using the \textit{splot} task in IRAF
\footnote{IRAF is distributed by the National Optical Astronomy Observatories, which are operated by the Association of Universities for Research in Astronomy, Inc., under cooperative agreement with the National Science Foundation.} 
and the method outlined by \cite{Looper08_dirty} of using measurements of multiple noise spikes to estimate upper limits and uncertainties.
Given that the \ion{Li}{1} line can be quite weak ($\sim$2~\AA), higher-resolution, higher signal-to-noise data are needed before the detection fraction of lithium in these low-gravity objects can be accurately measured. 

Unlike the other alkali metals, lithium can be fused in the interiors of brown dwarfs and cannot be replenished via nucleosynthesis.
As a result of this depletion over time, the strength of the line is sensitive to mass and age in addition to gravity.
%Lithium, unlike the other alkali metals, traces mass and age since
Objects less massive than $65~M_{Jupiter}$ are expected to retain most of their lithium and one would expect to see the \ion{Li}{1} absorption line in the spectra of young, low-mass brown dwarfs \citep{Burrows01}. 
However, most of our low-gravity objects fail the so-called ``lithium test'' \citep{Rebolo92}---lithium absorption is present in only nine out of twenty-three objects.
First of all, the strength of the line is expected to be weak at early types and increase towards the mid-to-late L dwarfs, as we observe on our objects.
Additionally, the \ion{Li}{1} line is probably weakened due to surface-gravity effects---just as the other alkali metals---rather than elemental depletion via core fusion.
The inherent weakness of the line at early-L types combined with low-gravity further weakening the line has probably pushed it beyond the detection limit of our moderate-resolution, moderate signal-to-noise data.
Thus, as concluded by \citet{Kirkpatrick08}, the ``lithium test'' is likely not a valid method of confirming the substellar nature of young L dwarfs.
However, further observations are needed of the lithium line in both our low-gravity objects and confirmed members of young clusters to bolster this conclusion.

\subsection{H$\alpha$ emission}
H$\alpha$ emission at 6563~\AA\ has been detected in the spectra of four of the twenty-three low-gravity L dwarfs: 2M~0141$-$46 (0--10.5~\AA), 2M~0712$-$61 (35$\pm$10~\AA), 2M~1022+58 (30--190~\AA), and 2M~0045+16 (13.1$\pm$2.2~\AA).
These data are shown in Figure~\ref{fig:sequence1}.
In our 2005 Oct 11 spectrum of 2M~0141$-$46, H$\alpha$ emission is not detected, however \cite{Kirkpatrick06} found emission present and variable in their spectra obtained on 2003 Dec 23 and 24.
As described by \cite{Schmidt07}, 2M~1022+58 exhibits strong and variable emission. 
Only one spectrum has been obtained of the other two objects so the time variability of their H$\alpha$ emission is unknown.

If these objects are indeed young as we suspect, the lack of H$\alpha$ emission in 19/23 objects is surprising.
The study of 1--5~Myr-old late-M dwarfs has shown H$\alpha$ emission to be pervasive and is often attributable to active accretion. 
\cite{West08} find the activity lifetimes of M5--M8 dwarfs to persist for 7--8~Gyr.
In the field, H$\alpha$ emission is not rare in early-L dwarfs and might even be common \citep{Schmidt07}. 
While there have been several measurements of the activity fraction of L dwarfs in the field \citep{NN,Schmidt07}, the age-activity relation for L dwarfs has not yet been studied.
Unfortunately, the \textit{lack} of detected H$\alpha$ emission is not a useful diagnostic since its absence does not rule out the presence of low-levels of accretion or strong magnetic fields.
Further multi-wavelength data and monitoring studies of L dwarfs of all ages are warranted to understand the nature of their magnetic fields and to determine if an age-activity relation exists for brown dwarfs.

\subsection{Photometric Properties: Red $J-K_s$ color}
\begin{figure}
\plotone{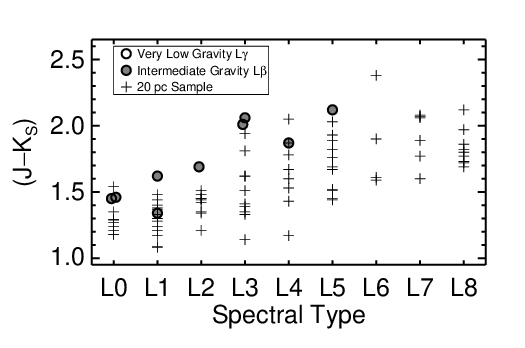}
\caption[$J-K_s$ color as a function of spectral type for the 20-pc sample and low-gravity L dwarfs.]{
	$J-K_s$ color as a function of spectral type.
	The colors of the very low-gravity (\textit{open circles}) and intermediate-gravity (\textit{shaded circles}) L dwarfs are compared to the colors of the objects in 20-pc sample (\textit{plusses}, \citealt{Paper10}). 
	The $J-K_s$ values for the low-gravity objects are listed in Table~\ref{tab:lowg}. 
	Overlapping low-gravity data points are slightly offset along the x-axis for clarity.}
\label{fig:color}
\end{figure}

In addition to having spectral peculiarities, low-gravity L dwarfs have red near-infrared colors.
The $J-K_s$ values are listed in Table~\ref{tab:lowg} and in Figure~\ref{fig:color} we compare the $J-K_s$ colors of the low-gravity objects to those of the 20-pc sample \citep{Paper10}.
The low-gravity objects fall at the red end of the distribution within their spectral class.
The latest, lowest-gravity object among the 23 low-gravity L dwarfs is also the reddest L dwarf found in the field to date: 2M~0355+11 has $J-K_s=2.52\pm0.03$ and we type it as a very low-gravity L5$\gamma$.

The second reddest objects shown in Figure~\ref{fig:color} is \object{2MASS J21481628+4003593}, which is in the 20-pc sample \cite[L6, $J-K_s$=2.38$\pm$0.04, ][]{Looper08_dirty}.
However, its spectrum does not display low-gravity features and \citeauthor{Looper08_dirty} conclude that its peculiarities are more likely due to unusually thick clouds rather than low gravity.
\textit{Thus, while all of the candidate low-gravity L dwarfs have quite red $J-K_s$ color, not all red L dwarfs necessarily have low surface gravities.}
Similarly, not all L dwarfs with blue near-infrared colors are necessarily old and metal poor: unresolved multiplicity and thin and/or large-grained condensate clouds can cause blue colors in addition to low metallicity and high gravity \citep{Cruz04,Burgasser08_0320, Burgasser08_blue}.

\subsection{Other Possible Explanations for Observed Features}

In the previous subsections, we described the peculiar features observed in our spectra and discussed their gravity dependence. In this section, we briefly address other effects that could be causing the observed spectral peculiarities.

\textit{Unresolved Binarity.}---
Spectra with unusual features can arise from the combined light from two unresolved sources with spectral types differing by two or more spectral types. 
This effect has been observed in the unresolved near-infrared spectra of binary MLT dwarfs \citep{Cruz04,Burgasser08_0320}.
However, in the case of L dwarfs, particularly in the optical, the light from the primary overwhelms the secondary due to the large brightness difference between spectral types \citep[see Figure 2 of ][]{Dahn02}. 
As a result, the combined spectrum is dominated by the spectrum of the primary while the secondary spectrum causes very subtle, if any, changes to it. 
To further explore this, we have experimented with making composite spectra from various combinations of normal L dwarf spectra. 
These combinations mirror the alkali lines and VO absorption of the primary spectrum and do not reproduce the deep VO absorption, sharper \ion{K}{1} lines, and weaker \ion{Na}{1} features that are present in the unusual spectra we present here. 
We therefore conclude that the peculiar features in our spectra are not due to unresolved binarity.

\textit{Metallicity.}---
\label{sec:metal}
Metal-poor L subdwarfs are characterized by weakened absorption bands of VO, stronger metal-hydride absorption, and a broader \ion{K}{1} doublet \citep{Burgasser07_subdwarfs}. 
These effects are nearly in the exact opposite sense of the unusual spectral features we observe in our peculiar objects and thus we conclude that low metallicity could not cause them.
On the other hand, one might argue that high photospheric metallicity could produce the stronger metal-oxide absorption bands and weakened metal-hydride bands that we observe.
The apparent pressure effects (e.g., sharp \ion{K}{1} lines) could be explained by reduced photospheric pressure in a high-metallicity environment instead of low gravity: $P \propto g/\kappa$, where $P$ is pressure, $g$ is gravity, and $\kappa$ is the mean opacity which is roughly proportional to metallicity \citep{Freedman08}.
However, our objects display weak TiO absorption bands, not stronger ones as is expected for high-metallicity dwarfs.
More importantly, very low metallicities ([M/H]$\sim-1$ to $-2$~dex) are required to alter the spectra of M and L subdwarfs significantly enough that they stand out from the spectra of normal dwarfs \citep{Gizis97, Burgasser07_subdwarfs}.
Hence, because it is expected that similarly extreme abundances would be needed to create noticeable spectral changes in high metallicity objects, and the most metal-rich stars in the Solar Neighborhood only have [M/H]$\lesssim +0.3$ \citep{Valenti05}, it is highly improbable that the peculiar spectral features we observe are due to exceptionally high metallicities.

\textit{Rotation.}---
High-resolution spectroscopy has shown the rotational velocities ($v_{rot}$~sin $i$) of normal field L dwarfs to range from 10 to 60~${\rm km~s^{-1}}$ \citep{Mohanty03,ZapateroOsorio07_rv}.
In general, fast rotation broadens both absorption lines and bands but does not alter their equivalent width.
Since we observe both narrower and weaker alkali absorption lines, rapid rotation could not be the culprit.
Neither slow nor fast rotation could cause deeper VO absorption.
Indeed, the spectral differences between the slow ($\sim 10~{\rm km~s^{-1}}$) and rapid rotators ($\sim 60~{\rm km~s^{-1}}$ ) are only perceptible at high resolution ($R>20,000$) and any rotation effects are not likely to be resolved in our modest $R\sim1000$ data.
Therefore, we conclude that rotation is not responsible for the spectral peculiarities observed in our spectra.

\textit{Clouds.}--- 
Condensate clouds are known to be present in the photospheres of L dwarfs and play a significant role in shaping their spectra.
The complex processes of condensate cloud formation and dynamics remains a topic of ongoing theoretical investigation \citep[see recent review by ][]{Helling08_compare}.
However, it appears that condensate opacity is mostly relevant in the near-infrared, where thicker clouds (dustier photospheres) give rise to redder near-infrared colors \citep[e.g., ][]{Ackerman01}.
%This affect is due to the suppression of the $J$ and $H$-band peaks where competing gas opacity is minimized (e.g., Ackerman & Marley 2001).  
%The redder near-infrared colors of dusty L dwarfs can potentially mimic the effects of reduced H$_2$ opacity in low-gravity L dwarfs.  
However, condensates are not expected to significantly influence optical spectra due to the greater gas opacities in this region, in particular the pressure-broadened \ion{K}{1} and \ion{Na}{1} lines \citep{Burrows03_K}.
This trend is confirmed by observed variations in the near-infrared colors and spectra of L dwarfs whose optical spectra are otherwise identical \citep{McLean03, Looper08_dirty, Burgasser08_blue}.
In addition, as VO is one of the condensates present in dusty L dwarf atmospheres, one would expect less VO gas absorption in the redder (dustier) L dwarfs, whereas we observer stronger VO absorption in our sources.
It has been suggested that thicker than normal clouds could be present in low-gravity L dwarf atmospheres \citep{Kirkpatrick06,Looper08_dirty}.
%just as kinematically older, higher-surface gravity objects appear to have reduced condensate abundances \citep{Burgasser08_blue, Faherty08}.  
Thus, while thicker condensate clouds may be present in the sources examined here, they do not appear to be responsible for the peculiar spectral features present at \textit{optical} wavelengths.

\subsection{Summary of Low-Gravity Spectral Features}

The spectra we present here deviate significantly from spectral standards and do not fit into the normal L dwarf sequence.
In particular, the peculiar spectra display weak \ion{Na}{1}, \ion{Cs}{1}, \ion{Rb}{1} lines; both weak line cores and weak pressure-broadened wings of \ion{K}{1}; weak FeH; weak TiO; and, at early types, strong VO.
We find that neither unresolved multiplicity, metallicity, rotation, nor clouds can explain these features. 
\textit{Based on the similarity of these features to those seen in the spectra of young (1~Myr) late-M dwarfs and late-M giants (both known to have low gravities) we conclude that the spectral peculiarities in our objects are caused by low gravity.}

\section{Discussion}
\label{sec:discussion}

\subsection{Ages}
\label{sec:ages}

Low gravity in brown dwarfs implies both low mass and youth.
However, it is difficult to estimate the ages of our low-gravity objects since the optical spectra of young L dwarfs in clusters are limited to early spectral types (L0--L1) and are not of sufficient signal-to-noise nor resolution to use as fiducials.
Despite the lack of an age-gravity calibration, we can still put rough age estimates on our low-gravity objects using their gravity-sensitive spectral features and their distribution on the sky.
 
As discussed by \citet[see their Fig.~7]{Kirkpatrick08}, only objects Pleiades age and younger ($\le$100~Myr) display low-gravity features discernible at the modest resolution of our data.
At these young ages, brown dwarfs are still contracting and have larger radii, and thus lower gravities, than older objects of the same temperature \citep{Burrows01}.
Since we can clearly see low-gravity effects in the spectra of our objects, we adopt 100~Myr as the upper limit on their age.
To estimate a lower age limit, we use the observation that none of our low-gravity objects are near young, dense star forming regions. 
Since the clusters these objects originated from appear to have dissipated, the objects are most likely $\sim$10~Myr or older.
Thus, we adopt an age range of $\sim$10--100~Myr for our low-gravity L dwarfs.

%"the strength is in the weakness"---LAM

As illustrated in the spectra shown in Figure~\ref{fig:sequence1} and~\ref{fig:sequence2} and quantitatively with the indices in Figures~\ref{fig:lines} and~\ref{fig:bands}, the low-gravity features in objects of the same spectral subtype have a range of strengths.
The spectra are shown from top-to-bottom within each spectral subtype in decreasing order of their spectral peculiarity with the normal-gravity standard shown last.
On the spectral index plots, objects with spectra indicative of very low-gravity are plotted as open circles while objects with intermediate-gravity features are plotted as shaded circles.
The weakness of the \ion{Na}{1} doublet and the \ion{K}{1} doublet's cores and wings are the primary features used to classify objects as either very low-gravity or intermediate-gravity.
As weaker alkali-metal features indicate lower gravities, the very low-gravity objects are also likely younger than the intermediate-gravity objects.
Thus we hypothesize that within each subtype, ordering the spectra based on their implied gravity also creates an age sequence.
Additionally, as long as the the effective temperatures of the younger objects are not significantly hotter than those of the older objects of the same spectral type, this sequence in gravity and age is also a mass sequence, with the youngest objects being the least massive.

\citet{Kirkpatrick08} use the behavior of low-gravity features in late-M dwarfs of different ages (see their Fig.~7) to demonstrate that gravity-sensitive features distinguish ages to $\approx$1~dex and we expect this remains true for L dwarfs.
For example, among the L0 objects, the low-gravity features suggest that 2M~0141$-$46 and its clones (L0$\gamma$) are the youngest with ages $\approx$10--30~Myr; 2M~1552+29 and DENIS~0357$-$44 (L0$\beta$) have intermediate ages of $\approx$100~Myr; and the L0 standard 2M~0345+25 is the oldest with an age $>$200~Myr.
Generalizing this idea implies that the objects we have classified as very low-gravity ($\gamma$) are closer to $\approx$10~Myr while the intermediate-gravity objects ($\beta$) are more likely $\approx$100~Myr.
The validity of this hypothesis for L dwarfs will likely not be determined until cluster membership is confirmed for our objects.
On the other hand, an age-gravity calibration for M dwarfs is possible and is underway and will likely shed light on the properties of low-gravity early-L dwarfs (Kirkpatrick et al., in preparation). 

\subsection{Distances/Temperatures}
\label{sec:distances}

We have estimated crude distances for the low-gravity objects with the $M_J$-spectral type relation derived by \citet{Cruz03}.
The uncertainties on the distances were estimated by propagating through the uncertainty on $M_J$ found for a spectral type uncertainty of $\pm$~1 and using the covariance matrix of the best fit parameters of the $M_J$-spectral type relation.
Field dwarfs with measured parallaxes listed by \cite{Dahn02} were used to derive this relation and it is not strictly valid for low-gravity objects. 
The $M_J$-spectral type relation has been updated to include new results and revised to include T dwarfs \citep[most recently by ][]{Looper08_T}. 
However, we continue to use the \citeauthor{Cruz03} relation to maintain consistency within our sample of objects found by the NStars 20-pc survey. 
The difference between the $M_J$ predicted by the two relations is at most 0.17~mag for L6 but is less than 0.1~mag for L4 and earlier.

Evaluating if the field $M_J$-spectral type relation under- or over-estimates the true distances to low-gravity objects is not straightforward. 
On one hand, the low-gravity objects could be brighter than normal objects of the same spectral type and the $M_J$-spectral type relation would underestimate luminosities and distances for low-gravity objects. 
This is indeed the case for higher-mass stars where the low-gravity objects (i.e., giants) have the similar effective temperatures to their normal gravity counterparts (i.e., dwarfs), but much larger radii.
This scenario would also be valid if the temperature scale for low-gravity objects is hotter than that of normal dwarfs as derived by \cite{Luhman99_IC348} and \cite{Luhman03_IC348}.

There are four objects with independently derived distances that might shed light on the validity of the $M_J$-spectral type relation for young objects. 
\citet{Kirkpatrick06} studied the low-gravity early-L dwarf 2M~0141$-$46 in detail and compared its observed optical and near-infrared spectra to model spectra; based on this thorough analysis, they estimate a distance of $\sim$35~pc. 
Using a spectral type of L0$\pm1$ for 2M~0141$-$46, the $M_J$-spectral type relation yields a distance of 41$\pm5$~pc. 
Similarly, \cite{Rebolo98} estimate a distance to G196-3A of $21\pm6$ and the $M_J$-spectral type relation gives $27\pm5$ using L3$\pm1$ for the low-gravity secondary.
In both cases, the two values agree within the uncertainties. 
On the other hand, \citet{Gizis07} and \cite{Teixeira08} find trigonometric parallax distances of $54.0^{+3.2}_{-2.8}$~pc and 55~pc for 2M~1207$-$39 and SSSPM~1102$-$34, respectively. 
Both of these objects are late-type M dwarfs in the TW Hydrae Association.
The $M_J$-spectral type relation yields a distance of 25$\pm$2 and 22$\pm$2 to the two systems respectively, an underestimate by a factor of two.
Each of these four systems are not ideal for evaluating the $M_J$-spectral type relation for young objects: the distance estimate for 2M~0141$-$46 is based on un-tested models and is not robust, the G196-3AB system might be older than our typical low-gravity L dwarf, and both SSSPM~1102$-$34 and 2M~1207$-$39 harbor active accretion disks \citep{Morrow08,Riaz08}.
Clearly, until trigonometric parallaxes are obtained for a relatively large sample of objects, both the temperature and distance scales will remain uncertain for low-gravity L dwarfs.

Regardless of the complications, the $M_J$-spectral type relation gives a rough estimate of the distance and at least provides a sense of relative distances to the low-gravity objects. 
For the lack of a better method, we adopt the distances predicted by the $M_J$-spectral type relation (with an uncertainty of $\pm$1 on the spectral type) for the low-gravity objects and list them in Table~\ref{tab:lowg}. 

\subsection{Distribution on the Sky and Coincidence with Moving Groups}

\begin{figure}
\plotone{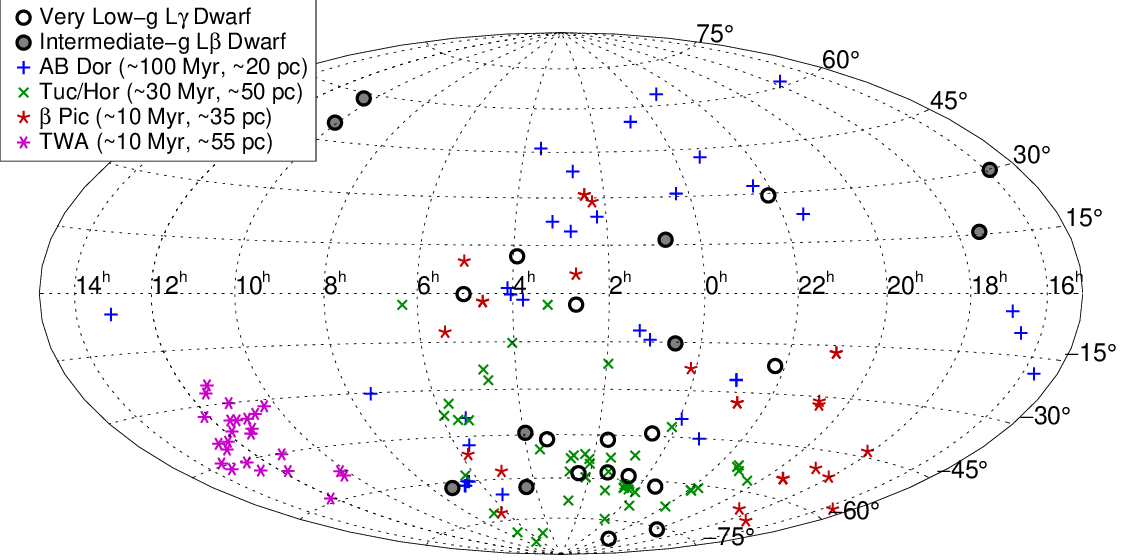}
\caption[Celestial distribution of young brown dwarf candidates and young stellar associations on the sky.]{
	Distribution on the sky in equatorial coordinates of the very low-gravity (\textit{open circles}) and intermediate-gravity (\textit{shaded circles}) L dwarfs. 
	Also shown are members of the AB Doradus moving group (\textit{blue pluses}), the Tucana/Horologium Association (\textit{green crosses}), the $\beta$ Pictoris moving group (\textit{red five-pointed asterisks}), and the TW Hydrae association (\textit{purple six-pointed asterisks}) as identified by \citet{Zuckerman04}. 
}
\label{fig:map}
\end{figure}

The location of our candidate low-gravity brown dwarfs on the sky is shown in Figure~\ref{fig:map} (\textit{open and shaded circles}). Also shown are members of the AB Doradus moving group ($\sim$100~Myr, \textit{blue plusses}), the Tucana/Horologium Association (Tuc/Hor, $\sim$30~Myr, \textit{green crosses}), the $\beta$ Pictoris moving group ($\sim$10~Myr, \textit{red five-point asterisks}), and the TW Hydrae association (TWA, $\sim$10~Myr, \textit{purple six-point asterisks}) as identified by \citet{Zuckerman04}. 
The positional coincidence and agreement between the estimated distances and ages all strongly suggest that our low-gravity L dwarfs are members of these groups.
Six of the nine very low-gravity L0$\gamma$ objects (\textit{open circles}) overlap the core of the Tuc/Hor association.
The intermediate-gravity dwarfs (\textit{shaded circles}) are more widely distributed on the sky and appear to have a distribution most similar the members of the AB Dor moving group. 
G196-3B and 2M~1022+58 are the outliers on the figure at 10$^{\mbox{h}}$, +50$\degr$ and do not not appear to be connected to any of the young associations considered here. 
This figure is similar to one presented in \citet{Kirkpatrick08} but here we show only L dwarfs and do not consider late-M dwarfs.
(As mentioned previously, the sample of twenty-three L dwarfs considered here and the twenty objects studied by \citet{Kirkpatrick08} have seven L dwarfs in common.)

The majority of the candidate low-gravity objects were identified via a color-magnitude query of the 2MASS All-Sky survey and thus their spatial distribution is not a selection effect.
By far, the largest position-based cuts in our 2MASS search were to avoid the crowded and reddened regions of the Galactic plane ($|b|>15\degr$) and known star formation regions that extend beyond the plane (e.g., Upper Sco, Taurus).
Other small regions were also excluded such as M31, M33, and the Large and Small Magellanic Clouds.
As shown in Figure~1 of \cite{Paper10}, the distribution of the sources surviving these cuts, which is the parent population of our low-gravity objects, is uniform.
Again, the selection criteria and sky coverage of our 2MASS search are discussed in detail by \cite{Cruz03, Cruz07} and \citeauthor{Paper10}.

In addition to the coincidence of the projected location on the sky of our low-gravity objects with the moving groups, their distance estimates are also broadly consistent with each other.
The M$_J$-spectral type relation yields distances to the low-gravity objects all within 85~pc; the majority are estimated to be between 13~and 50~pc.  
(As described in the previous section, these distances are very uncertain.)
The mean distances to the moving groups are also between 20 and 50~pc; however, the spread of distance estimates within each group is quite large. 
For example, as listed by \citet{Zuckerman04}, the distances to $\beta$~Pic members range from 15--60~pc and 37--66~pc for Tuc/Hor members.
As such, unlike most of their young cluster counterparts, these nearby moving groups overlap with each other and have neither a unique distance nor unique location on the sky.
Thus, accurate distances alone for the low-gravity objects would not determine which group they belong to.
For these reasons, we are currently targeting all of our low-gravity objects for proper motion, radial velocity, and parallax measurements in order to derive their UVW space motions and to confirm or refute cluster membership \citep{Mamajek05}.

None of the low-gravity L dwarfs appear to be members of TWA.
Indeed, the only TWA member uncovered by the 20-pc census is \object[2MASS J11395113-3159214]{2MASS J11395113$-$3159214} (M9).
The other late-M dwarf members were excluded by the color cuts: \object[2MASS J12073346-3932539]{2MASS J12073346$-$3932539} failed the $J-K_s$ cut while \object[2MASS J11020983-3430355]{2MASS J11020983$-$3430355} failed the $R-J$ cut \citep{Cruz03}.
With a mean distance of $\sim$54~pc, TWA is the most distant of the nearby associations and any L dwarf members were probably also excluded by the color-magnitude search criteria designed to constrain the sample to only the nearest/brightest and latest-type/reddest dwarfs.
While recent work using the Deep Near-Infrared Southern Sky Survey (DENIS) has identified one new late-type M dwarf member of TWA \citep{Looper07_1245}, no L dwarfs were found by this effort (Looper 2008, private communication). 
Despite its large spatial extent, a deep survey of TWA is warranted to uncover its later-type members.

\subsection{Masses}
Evolutionary models, combined with reasonable assumptions about ages and temperatures, can be used to estimate the masses of the low-gravity L dwarfs. 
The temperature scale for field L dwarfs has been found to range from 2400~K at L0 down to 1500~K for L5--L8 \citep{Golimowski04}.
Even though the temperature scale for low-gravity L dwarfs is likely to be different than that of normal-gravity field dwarfs, it is sufficient to make the rough mass estimates we are interested in here.
For field-age (1--5 Gyr) objects within that temperature range, the evolutionary models of \cite{Burrows01} and \cite{Baraffe03} predict masses of 48--85~$M_{Jupiter}$.
If, however, the low-gravity L dwarfs are indeed young as we suspect, the models predict very-low masses: at 30~Myr, the age of Tuc/Hor, 1500--2400~K corresponds to 11--30~$M_{Jupiter}$.

One individual object, 2M~0355+11, merits comment.
We characterize this object's peculiar spectrum as a very low-gravity L5$\gamma$ (Figure~\ref{fig:sequence2}).
This interpretation puts this object at the lowest end of both our age and temperature estimates and we speculate it has on effective temperature of 1500--1700~K (based on its similarity to an L5) at an age of 10--30~Myr (based on its very  low-gravity spectral features).
In this scenario, the evolutionary models predict a mass of 8--13~$M_{Jupiter}$.
If indeed this object is quite young and very low-mass, it would be the lowest-mass L dwarf identified in the field and an analog to 2M~1207$-$39B, the very cool L dwarf companion found in TWA \citep{Chauvin05}.
Both of these objects have very red $J-K$ colors and are the two reddest L dwarfs known: 2M~0355+11 and 2M~1207$-$39B have $J-K_s$ colors of 2.52$\pm$0.03 and 3.07$\pm$0.23 respectively \citep{Mohanty07}.
However, further age and effective temperature constraints are necessary before a robust mass estimate for 2M~0355+11 can be made.

%Teff=3759-135x (L0: x=10). Age=10-100Myr.
%L0=2400 K
%L5=1734

\section{Summary}
\label{sec:summary}

We have identified 23 L dwarfs with very peculiar spectral features in their optical spectra. 
The peculiar features that we observe include weak alkali doublets (\ion{Na}{1}, \ion{K}{1}, \ion{Rb}{1}, \ion{Cs}{1}); weak \ion{K}{1} wings; weak absorption bands of FeH, CrH, and TiO; and strong VO absorption bands at early types.
All of these spectral peculiarities, except weak TiO, are consistent with the spectral features characterizing low-gravity objects (i.e., giants and young M dwarfs).
After considering and ruling out other possible explanations for the observed spectral peculiarities, we conclude low-gravity is the most probable underlying cause.
We use these spectra to expand the \cite{K99} scheme to include three gravity classes spanning L0 to L5.
The gravity of the spectrum is indicated by a greek suffix: $\alpha$ (or lack of suffix) for normal gravity, $\beta$ for intermediate-gravity, and $\gamma$ for very-low gravity.
We do not propose any modification to the \cite{K99} method for assigning spectral types to normal-gravity L dwarfs.
Compared to normal dwarfs with the same spectral type, the low-gravity objects have red $J-K_s$ colors.
Contrary to conventional wisdom, only 9/23 display \ion{Li}{1} in absorption and only 3/23 objects display H$\alpha$ in emission. 
However, higher signal-to-noise observations are required to robustly measure the detection fraction of these lines in low-gravity L dwarfs.

Based on the low-gravity features present in the spectra of our objects and their lack of association with young, dense star-forming regions, we estimate their ages to be approximately 10--100~Myr.
Further, given the behavior of low-gravity features in the spectra of late-M dwarfs of different ages shown by \citet{Kirkpatrick08}, we assert that the objects we have classified as very low gravity are closer to $\approx$10~Myr while the intermediate-gravity objects are more likely $\approx$100~Myr.
While the temperature and distance scales for low-gravity L dwarfs are still uncertain, we estimate rough distances to our objects using the \citet{Cruz03} $M_J$-spectral type relation for field dwarfs.
There is positional, age, and distance agreement between the low-gravity L dwarfs and southern, nearby loose associations, most notably Tuc/Hor.
However, before membership in any association or moving group can be assigned and ages adopted, astrometric and radial velocity measurements are required and those observations are underway. 
Future statistically robust studies of these nearby associations could yield an initial mass function that probes the planetary-mass regime and possibly identify the true lowest-mass objects formed via star formation.

\acknowledgments
We are very grateful for useful discussions with Michael Cushing, Greg Herczeg, Lynne Hillenbrand, Adam Kraus, Dagny Looper, Jessica Lu, Lauren MacArthur, Mark Marley, and I.~Neill Reid.
We also thank the referee for helpful comments.
K.~L.~C. is supported by NASA through the Spitzer Space Telescope Fellowship Program, through a contract issued by the Jet Propulsion Laboratory, California Institute of Technology under a contract with National Aeronautics and Space Administration.
This research was partially supported by a grant from the NASA/NSF NStars initiative, administered by the Jet Propulsion Laboratory, Pasadena, CA. 
This paper includes data gathered with the 6.5~m Magellan Telescopes located at Las Campanas Observatory, Chile.
This publication makes use of data products from the Two Micron All Sky Survey, which is a joint project of the University of Massachusetts and Infrared Processing and Analysis Center/California Institute of Technology, funded by the National Aeronautics and Space Administration and the National Science Foundation.
This research has made use of the NASA/IPAC Infrared Science Archive, which is operated by the Jet Propulsion Laboratory/California Institute of Technology, under contract with the National Aeronautics and Space Administration.

\bibliographystyle{apj}
\bibliography{bib_all}

\begin{deluxetable}{llllllcccc}
\tabletypesize{\tiny}
\rotate
\tablewidth{0pt}
\tablecolumns{10}
\tablecaption{Low-Gravity L Dwarfs \label{tab:lowg}}
\tablehead{ 
\colhead{2MASS} & 
\colhead{} & \colhead{} & \colhead{} &
\colhead{Obs. Date} & \colhead{} & 
\colhead{Optical} & \colhead{\ion{Li}{1} EW} &
\colhead{estimated} &
\colhead{} \\
\colhead{Designation\tablenotemark{a}}  &
\colhead{$J$} & \colhead{$J-H$} & \colhead{$J-K_s$} & 
\colhead{(UT)} & \colhead{Telescope} &
\colhead{Spec. Type\tablenotemark{b}} & \colhead{(\AA)} &
\colhead{d (pc)} &
\colhead{Refs.}
}
\startdata
\cutinhead{Very Low-Gravity L0$\gamma$}
00325584$-$4405058 &  $14.78\pm0.04$ & $0.92\pm0.05$ & $1.51\pm0.05$ & 2006 Jan 15    & CT 4m-RC    & L0$\gamma$  & $<5$        & $41\pm5$  & 1    \\ %paper10
00374306$-$5846229 &  $15.37\pm0.05$ & $1.12\pm0.07$ & $1.79\pm0.07$ & 2005 Aug 12/19 & GS-GMOS     & L0$\gamma$  & $<2$        & $54\pm7$  & 1    \\ %paper10 \\
01244599$-$5745379 &  $16.31\pm0.11$ & $1.25\pm0.14$ & $1.99\pm0.14$ & 2005 Aug 19    & GS-GMOS     & L0$\gamma$  & $<3$        & $82\pm11$ & 1    \\ %paper10 \\
01415823$-$4633574 &  $14.83\pm0.04$ & $0.96\pm0.05$ & $1.74\pm0.05$ & 2005 Oct 11    & GS-GMOS     & L0$\gamma$  & 1.3$\pm$0.5 & $42\pm5$  & 1--3 \\%kirkpatrick06 \\
02235464$-$5815067 &  $15.07\pm0.05$ & $1.07\pm0.06$ & $1.65\pm0.07$ & 2005 Aug 19    & GS-GMOS     & L0$\gamma$  & $<1$        & $47\pm6$  & 1    \\%paper10 \\
02411151$-$0326587 &  $15.80\pm0.07$ & $0.99\pm0.08$ & $1.76\pm0.08$ & 2005 Oct 10    & GS-GMOS     & L0$\gamma$  & $<2$        & $66\pm8$  & 3, 4    \\ %cruz07 \\
03231002$-$4631237 &  $15.39\pm0.07$ & $1.07\pm0.09$ & $1.69\pm0.09$ & 2005 Jan 01    & GS-GMOS     & L0$\gamma$  & 2.3$\pm$1.4 & $54\pm7$  & 1 \\%paper10\\
19564700$-$7542270 &  $16.15\pm0.10$ & $1.12\pm0.14$ & $1.93\pm0.12$ & 2005 Aug 10    & GS-GMOS     & L0$\gamma$  & $<2$        & $77\pm10$ & 1 \\%paper10 \\
22134491$-$2136079 &  $15.38\pm0.04$ & $0.97\pm0.07$ & $1.62\pm0.05$ & 2005 Sep 08    & GS-GMOS     & L0$\gamma$  & $<1$        & $54\pm7$  & 3, 4 \\ %cruz07 \\
\cutinhead{Intermediate-Gravity L0$\beta$}
03572695$-$4417305AB\tablenotemark{c} & $14.37\pm0.03$ & $0.84\pm0.04$ & $1.46\pm0.04$ & 2003 Dec 24 & Keck-LRIS & L0$\beta$ & $<2$ & $46\pm3$ & 3--6 \\%davyLs,bouy03,martin06 \\
15525906+2948485   &  $13.48\pm0.03$ & $0.87\pm0.04$ & $1.46\pm0.04$ & 2004 Feb 02    & KP 4m-MARS  & L0$\beta$   & $<2$        & $22\pm3$  & 1, 7, 8\\%wilson03, jameson08, paper10\\
\cutinhead{L1}                                                     
07123786$-$6155528\tablenotemark{d} &  $15.30\pm0.06$ & $0.90\pm0.08$ & $1.62\pm0.08$ & 2008 Feb 22  & Clay-LDSS3 & L1$\beta$ & $<3$ & $46\pm6$  & 9 \\ %burgasser07
10224821+5825453   &  $13.50\pm0.03$ & $0.86\pm0.04$ & $1.34\pm0.04$ & 2004 Feb 12    & KP 4m-MARS  & L1$\beta$  & $<1$        & $20\pm3$  & 1, 4, 10 \\%Paper 10, Schmidt 2007
\cutinhead{L2}
00452143+1634446   &  $13.06\pm0.02$ & $1.00\pm0.04$ & $1.69\pm0.03$ & 2003 Jul 10    & KP 4m-MARS  & L2$\beta$   & $<3$        & $14\pm2$  & 1, 7\\%wilson03, paper 10. has h alpha.\\
23225299$-$6151275 &  $15.55\pm0.06$ & $1.01\pm0.09$ & $1.69\pm0.08$ & 2005 Aug 12    & GS-GMOS     & L2$\gamma$  & $<2$        & $44\pm7$  & 1 \\%paper10 \\
\cutinhead{L3}
10042066+5022596\tablenotemark{d} & $14.83\pm0.05$ & $1.18\pm0.06$ & $2.05\pm0.06$ & 1999 Mar 04/05 & Keck-LRIS & L3$\beta$ & 5.5$\pm$0.7 & $27\pm5$ & 11--13\\%Rebolo98, K01-gstar, McGovern04  \\
....aka G196-3B    &  &  &                & 2001 Feb 19    & Keck-LRIS   &       &             &           & \\
17260007+1538190   &  $15.67\pm0.07$ & $1.20\pm0.08$ & $2.01\pm0.08$ & 1998 Aug 12    & Keck-LRIS   & L3$\beta$   & 4.4$\pm$1.6 & $39\pm7$  & 14 \\%K00
21265040$-$8140293 &  $15.54\pm0.06$ & $1.14\pm0.08$ & $2.00\pm0.07$ & 2004 Dec 09    & GS-GMOS     & L3$\gamma$  & 4.2$\pm$1.8 & $37\pm7$  & 1 \\%paper10 \\
22081363+2921215   &  $15.80\pm0.09$ & $1.00\pm0.11$ & $1.65\pm0.11$ & 1998 Dec 25    & Keck-LRIS   & L3$\gamma$  & 3.0$\pm$1.9 & $42\pm8$  & 4, 14 \\%K00 \\
\cutinhead{L4}
00332386$-$1521309 &  $15.29\pm0.06$ & $1.08\pm0.08$ & $1.88\pm0.07$ & 2005 Oct 10    & GS-GMOS     & L4$\beta$   & $<1$        & $27\pm6$  & 1, 3, 4, 15\\%paper10, davyls, gizis03. \\
05012406$-$0010452 &  $14.98\pm0.04$ & $1.27\pm0.05$ & $2.02\pm0.05$ & 2005 Oct 10    & GS-GMOS     & L4$\gamma$  & 9.0$\pm$2.4 & $24\pm5$  & 1 \\%paper10 \\
\cutinhead{L5}
03552337+1133437   &  $14.05\pm0.02$ & $1.52\pm0.04$ & $2.52\pm0.03$ & 2005 Nov 27    & GN-GMOS     & L5$\gamma$  & 6.9$\pm$0.8 &  $13\pm3$ & 1, 16 \\%Lithium. Blake07, paper10 \\
04210718$-$6306022 &  $15.57\pm0.05$ & $1.28\pm0.06$ & $2.12\pm0.07$ & 2004 Dec 12    & GS-GMOS     & L5$\beta$   & 7.2$\pm$1.7 &  $25\pm6$ & 4 \\%Lithium. Cruz07 \\
\enddata

\tablecomments{
Objects are grouped by spectral type and then listed in order of right ascension. 
The first group of very low-gravity L0$\gamma$ dwarfs includes 2M~0141$-$46 \citep{Kirkpatrick06} and eight objects with spectra very similar to it as shown in Fig.~\ref{fig:clones}. 
The two intermediate-gravity L0$\beta$-type objects and the eleven later-type objects (L1--L5) are shown in Figs.~\ref{fig:sequence1} and ~\ref{fig:sequence2}.
In Figures~\ref{fig:lines}--\ref{fig:map}, open and shaded circles are used to distinguish the very low-gravity (``$\gamma$'') and intermediate-gravity (``$\beta$'') objects, respectively.
For several objects, optical spectra obtained with different setups exist; here we provide references for other data and only list the observations relevant to the data used in this paper.}

\tablenotetext{a}{The sexagesimal right ascension and declination suffix of the full 2MASS All-Sky Data Release designation 
(2MASS Jhhmmss[.]$\pm$ssddmmss[.]s) is listed for each object.
The coordinates are given for the J2000.0 equinox; the units of right ascension are hours, minutes, and seconds; 
and units of declination are degrees, arcminutes, and arcseconds.}

\tablenotetext{b}{
Spectral types are tentative and were assigned using an expanded version of the \cite{K99} spectral typing scheme for normal L dwarfs as described in \S~\ref{sec:types}.
Objects with intermediate low-gravity features are indicated with an ``$\beta$'' suffix while ``$\gamma$'' indicates objects with spectral features indicating very low-gravity. }

\tablenotetext{c}{\citet{Bouy03} resolve this object, aka DENIS~0357$-$44, into a binary system with $\Delta I \sim 1.5$. 
Low signal-to-noise resolved spectra obtained by \citet{Martin06} reveal the components to differ by at least two spectral types (e.g., M9 and L1).} 

\tablenotetext{d}{These spectra have been corrected for telluric absorption.}

\tablerefs{
(1)~\citet{Paper10};
(2)~\citet{Kirkpatrick06};
(3)~\citet{Kirkpatrick08};
(4)~\citet{Cruz07};
(5)~\citet{Bouy03};
(6)~\citet{Martin06};
(7)~\citet{Wilson03};
(8)~\citet{Jameson08};
(9)~\citet{Burgasser08_blue}
(10)~\citet{Schmidt07};
(11)~\citet{Rebolo98};
(12)~\citet{K01_gstar} 
(13)~\citet{McGovern04};
(14)~\citet{K00};
(15)~\citet{Gizis03}
(16)~\citet{Blake07};
}

\end{deluxetable}

\end{document}